\documentclass [12pt,a4paper] {article}

\usepackage [latin1] {inputenc}
\usepackage {graphicx}
\usepackage {url}

\usepackage [colorlinks=true,urlcolor=blue,citecolor=red,linkcolor=blue] {hyperref} 

\begin {document}
\pagestyle {plain}

\title {\bf Wilhelm Weber: On the Energy of Interaction (translated and edited by A. K. T. Assis)}
\author {\bf Wilhelm Weber}
\date {}
\maketitle

\vskip2cm

{\it Editor's Note: An English translation of Wilhelm Weber's 1878 paper ``Ueber die Energie der Wechselwirkung'', \cite {weber1878b}. This work is an excerpt from Weber's seventh major Memoir on {\it Electrodynamic Measurements}, ``Elektrodynamische Maassbestimmungen'', \cite {weber1878a} with English translation in \cite {weber2020c}.}
\vskip1cm
Third version posted in February 2021 (first version posted in September 2020) at \url {www.ifi.unicamp.br/~assis} and \url {https://arxiv.org/abs/2009.09296}

\vskip2cm

\pagestyle {myheadings}
\markboth {} {}

\newpage

By Wilhelm Weber\footnote {\cite {weber1878b}, related to \cite {weber1878a} with English translation in \cite {weber2020c}.}$^,$\footnote {Translated and edited by A. K. T. Assis, \url {www.ifi.unicamp.br/~assis}. I thank F. D. Tombe for relevant suggestions.}$^,$\footnote {The Notes by H. Weber, the Editor of Volume 4 of Weber's {\it Werke}, are represented by [Note by HW:]; while the Notes by A. K. T. Assis are represented by [Note by AKTA:].}

\vskip1cm

(Excerpt by the author from the Treatise on {\it Elek\-tro\-dy\-na\-mi\-sche Maass\-be\-stim\-mun\-gen} in Volume XVIII of the {\it Königl. Sächs. Gesellschaft der Wis\-sen\-schaf\-ten}.)\footnote {[Note by HW:] {\it Annalen der Physik und Chemie}, edited by G. Wiedemann, Vol. 4, Leipzig, 1878, pp. 343-373.}$^,$\footnote {[Note by HW:] As \S 1-5 of the excerpt coincides in content and wording with \S 1-5 of the previous treatise, in fact up to page 382 line 10 from above, only the last Section of the excerpt, \S 6, has been printed here.}$^,$\footnote {[Note by AKTA:] Pp. 343-365 of \cite {weber1878b} coincide with pp. 645-664 line 12 from above of the {\it Abhandlungen der mathematisch-physischen Classe der K\"oniglich S\"achsischen Gesellschaft der Wissenschaften (Leipzig)}, \cite {weber1878a}, and with pp. 364-382 line 10 from above of Volume 4 of Weber's {\it Werke}, \cite {weber1894a}.}

\vskip1cm

\centerline {\bf 6. A Particle Driven by both an Electric and a Non-Electric Force}
\centerline {\bf while Enclosed in an Electrified Spherical Shell}

\vskip.5cm

Regarding the applications of the fundamental electric law, in order to show that none of the ``inconsistent and absurd'' consequences occur, through which Helmholtz wished to refute this fundamental law, we will only consider here the application to the motion of a mass point $\mu$ (with an electric quantum $\varepsilon$) enclosed in an {\it electric spherical shell}, when acted on by both an {\it electric} force and a {\it non-electric} constant force $a$.\footnote {[Note by AKTA:] Weber is referring here to his electrodynamic force law which he presented in 1846, \cite {weber1846} with partial French translation in \cite {weber1887} and a complete English translation in \cite {Weber2007b}.

Weber studies in this paper of 1878 the motion of a particle with mass $\mu$ and electric charge $\varepsilon$ moving inside a uniformly electrified spherical shell. He considers two forces acting on this particle, namely, the electric force exerted by the shell and a non-electric constant force $a$. He considers this constant force $a$ to be the weight of the particle near the surface of the Earth, namely, $a = \mu g$. He is replying to Helmholtz's criticisms presented in 1873, \cite {helmholtz1873}, see also \cite {helmholtz1872b} with English translation in \cite {helmholtz1872a}.}

From this fundamental law, Helmholtz deduced in Borchardt's {\it Journal}, [Volume] LXXV,\footnote {[Note by AKTA:] \cite {helmholtz1873}; see also \cite {helmholtz1872b} with English translation in \cite {helmholtz1872a}.} the equation of the {\it vis viva} for this mass point $\mu$ with electric quantum $\varepsilon$, [inside] a spherical shell of radius $R$ uniformly covered with electricity, which appears as follows:\footnote {[Note by AKTA:] \cite [Section 12, pp. 48-54] {helmholtz1873}, see also \cite [\S\S 3 and 7] {neumann1874}.

The Latin expression {\it viv viva} ({\it living force} in English or {\it lebendige Kraft} in German) was coined by G. W. Leibniz (1646-1716). 

Originally the {\it vis viva} of a body of mass $m$ moving with velocity $v$ relative to an inertial frame of reference was defined as $mv^2$, that is, twice the modern kinetic energy. However, during the XIXth century many authors like Weber and Helmholtz defined the {\it vis viva} as $mv^2/2$, that is, the modern kinetic energy.

Weber, for instance, in his paper of 1871 on the conservation of energy discussed two electrified particles of charges $e$ and $e'$ separated by a distance $r$. He then said the following, \cite [Footnote 1, pp. 256-257 of Weber's {\it Werke}] {weber1871} with English translation in \cite [p. 9] {weber1872}:

\begin {quote}
If $\varepsilon$ and $\varepsilon'$ denote the masses of the particles $e$ and $e'$, and $\alpha$ and $\beta$ the velocities of $\varepsilon$ in the direction of $r$ and at right angles thereto, and $\alpha'$ and $\beta'$ the same velocities for $\varepsilon'$, so that $\alpha - \alpha' = dr/dt$ is the relative velocity of the two particles, then

\[
\frac {1}{2}\varepsilon\left(\alpha^2 + \beta^2\right) + \frac {1}{2}\varepsilon'\left(\alpha'^2 + \beta'^2\right)
\]
is the total {\it vis viva} of the two particles.
\end {quote}

In 1872 Helmholtz expressed himself as follows, \cite [p. 533] {helmholtz1872a}:

\begin {quote}
If we, as has always hitherto been done, name {\it vis viva} or {\it actual energy} the sum of the moved inert masses multiplied each by half the square of its velocity, then, [...]
\end {quote}}

\[
  \frac{1}{2}\left(\mu-\frac{8\pi}{3c^2}\cdot R \varepsilon
   \varepsilon^\prime\right) q^2
  -V+C=0\ ,
\]
where $\varepsilon^\prime$ denotes the quantum of electricity per unit area on the surface of the spherical shell, $q$ the velocity of the mass point
$\mu$ and $V$ the potential
of the \emph{non-electric} force.\footnote {[Note by AKTA:] $\varepsilon^\prime$ is the surface charge density. The total charge over the whole surface of the spherical shell of radius $R$ is then given by $4\pi R^2\varepsilon^\prime$.}

From this equation it has been concluded that when, with an existing {\it difference between the potential $V$ of the non-electric force and the constant} $C$, $\varepsilon'$ would have increased from 0 to $[8\pi/3c^2]R\varepsilon\cdot\varepsilon' = \mu$, then the {\it vis viva} of the point mass $\mu$ would have increased from $\frac {1}{2}\mu q^2 = V -\ C$ up to $\frac {1}{2}\mu q^2 = \infty$, which would be an {\it infinitely large work output}.\footnote {[Note by AKTA:] {\it Arbeitsleistung} in the original. This expression can also be translated as ``work performed''.} The removal of this objection can now be obtained from the complete presentation of the whole process of motion in its context, as indicated earlier in these {\it Annalen}, [Volume] XLVI, p. 29.\footnote {[Note by HW:] Wilhelm Weber's {\it Werke}, Vol. IV, p. 333.}$^,$\footnote {[Note by AKTA:] \cite [p. 29 of the {\it Annalen der Physik und Chemie} and p. 333 of Weber's {\it Werke}] {weber1875}.} 

Let us denote by $\eta$ that charge $\varepsilon'$ on the unit area of the spherical shell for which the velocity $q$ of the mass $\mu$ would be infinite, then set $\eta = [3c^2\mu/8\pi R\varepsilon$], and assume that $\varepsilon$ has a certain constant value, while $\varepsilon'$ grows uniformly from 0 at time $t = -\vartheta$ up to $\eta$ at time $t = 0$, the latter value being gradually attained. Furthermore, to simplify the analysis, take the center of the sphere as the starting point of the path $s$\footnote {[Note by AKTA:] Weber will consider the motion of the particle along a straight line beginning at the center of the shell. We can represent this motion as taking place along the $x$ axis, with $x = 0$ at the center of the shell, so that the path or trajectory $s = x$ might have positive or negative values. When $s = \pm R$ the particle would reach the spherical shell of radius $R$.} where the particle $\mu$ at time $t = -\vartheta$ (where $\varepsilon' = 0$) is at rest, that is, with $\varepsilon^\prime = 0$\footnote {[Note by AKTA:] Due to a misprint this expression appeared in the original as $\varepsilon = 0$.} we have $s = 0$ and $q = 0$. Then with the help of the values

\[
\varepsilon' = \eta\left(1 + \frac {t}{\vartheta}\right)\ ,\ \ \ \ \ \mu = \frac {8\pi}{3c^2}\cdot R\varepsilon\eta\ \ \ \ \ {\rm and}\ \ \ \ \ \frac {dV}{ds} = a\ ,
\]
(see Article 12 of the {\it Abhandlung})\footnote {[Note by HW:] Wilhelm Weber's {\it Werke}, Vol. IV, p. 333.}$^,$\footnote {[Note by AKTA:] \cite [p. 29 of the {\it Annalen der Physik und Chemie} and p. 333 of Weber's {\it Werke}] {weber1875}.}  the following equation is obtained:

\[
dq = - \frac {a\vartheta}{\mu}\cdot \frac {dt}{t}\ .
\]
The integral of this equation can be written as:\footnote {[Note by AKTA:] What Weber writes here as $\log$ of a magnitude $\theta$ should be understood as the natural logarithm of $\theta$ to the base of Euler's constant $e = 2.718...$, namely, $\log\theta = \log_e\theta = \ln \theta$. His integration can be expressed as follows:

\[
\int_{q = 0}^q dq = -\frac {a\vartheta}{\mu}\int_{t = -\vartheta}^t\frac {dt}{t} = -\frac {a\vartheta}{\mu}\left[\ln|t|\right]_{t = -\vartheta}^t = -\frac {a\vartheta}{\mu}\ln\sqrt{\frac {t^2}{\vartheta^2}}\ ,
\]
such that

\[
q = -\frac {a\vartheta}{2\mu}\ln\frac {t^2}{\vartheta^2}\ .
\]}

\[
q = -\frac {a\vartheta}{2\mu}\cdot\log C^2t^2\ ,
\]
in which $C^2 = 1/\vartheta^2$, because $q = 0$ should take place for $t = -\vartheta$. Therefore, as $q = ds/dt$:

\[
ds = -\frac {a\vartheta}{2\mu}\cdot\log\frac {t^2}{\vartheta^2}\cdot dt\ .
\]
From this it follows through integration:

\[
s = \frac {a\vartheta}{\mu}\left(1 - \frac {1}{2}\log\frac {t^2}{\vartheta^2}\right)\cdot t + C'\ .
\]
Since now $s = 0$ for $t = -\vartheta$, it results $C' = a\vartheta^2/\mu$, therefore:

\[
s = \frac {a\vartheta^2}{\mu}\left(1 + \frac {t}{\vartheta}\left(1 - \frac {1}{2}\log\frac {t^2}{\vartheta^2}\right)\right)\ .
\]

When we set the {\it non-electric} force acting on $\mu$ as $a = g\mu$, with $q'$ being the ratio of the velocity $q$ to $g\vartheta$, and with $s'$ being the ratio of $s$ [the path] to $g\vartheta^2$, then these formulas can be written as:\footnote {[Note by AKTA:] Weber is assuming here that the constant force $a$ is the weight of the particle of mass $\mu$ near the surface of the Earth, namely, $a = \mu g$. Moreover, he is defining the dimensionless displacement $s' = s/(g\vartheta^2)$ and the dimensionless velocity $q' = q/(g\vartheta) = (ds/dt)/(g\vartheta)$.}

\[
\frac {dq'}{dt} = -\frac {1}{t}\ ,
\]

\[
q' = -\frac {1}{2}\log\frac {t^2}{\vartheta^2}\ ,
\]

\[
s' = 1 + \frac {t}{\vartheta}\left(1 - \frac {1}{2}\log\frac {t^2}{\vartheta^2}\right)\ .
\]
Now they can be used for the construction of all motions of the particle $\mu$ with an uniformly growing charge $\varepsilon'$ and can be represented in a tabular overview, where $e$ is the base of the natural logarithm:\footnote {[Note by AKTA:] Instead of $dq'/dt$, the expression in the fourth column of the first line in the next Table should be the dimensionless acceleration given by 

\[
\frac {1}{g}\frac {d^2s}{dt^2} = \frac {1}{g}\frac {dq}{dt} = \vartheta\frac {dq'}{dt} = - \frac {\vartheta}{t}\ .
\]}

\begin{center}
\begin{tabular}{|c|c|c|c|c|}
\hline
$\frac{t}{\vartheta}$ & $s'$ & $q'$ & $\frac {dq'}{dt}$ & $\frac{\varepsilon'}{\eta}$
\\
\hline
$-1$&$0$&$0$ & +1 &$0$\\
$-e^{-1}$&$1-2e^{-1}$&$1$& $+e$ & $1-e^{-1}$\\
$-e^{-2}$&$1-3e^{-2}$&$2$& $+e^2$ & $1-e^{-2}$\\
$-e^{-3}$ & $1 - 4e^{-3}$ & 3 & $+e^3$ & $1 - e^{-3}$\\
\vdots&\vdots&\vdots&\vdots&\vdots\\
$0$&$1$&$\infty$& $\pm \infty$ & $1$\\
\vdots&\vdots&\vdots&\vdots&\vdots\\
$+ e^{-3}$ & $1 + 4e^{-3}$ & 3 & $-e^3$ & $1 + e^{-3}$\\
$+e^{-2}$&$1+3e^{-2}$&$2$& $-e^2$ & $1+e^{-2}$\\
$+e^{-1}$&$1+2e^{-1}$&$1$& $-e$ & $1+e^{-1}$\\
$+1$&$2$&$0$& $-1$ & $2$\\
$+e$&$1$&$-1$& $-e^{-1}$ & $1+e$\\
$+e^2$&$1-e^2$&$-2$& $-e^{-2}$ & $1+e^2$\\
\hline
\end {tabular}
\end {center}

The curve $ABCDEFGH$ in the next Figure represents, according to this information, the dependence of the velocity $q'$ as a function of the path length $s'$, namely, $s'$ as abscissa and $q'$ as ordinate. This curve goes from the center $A$ of the sphere as the starting point of the coordinates out to $B$, $C$ and approaches asymptotically the {\it ordinate} for $s' = 1$, then returning from there to $D$, $E$, $F$, where it intersects the axis of abscissas at the point $s' = 2$, and then goes on to $G$ and $H$, where $s$ becomes $= -R$ and $\mu$ hits the spherical shell.\footnote {\label {foot-Ho} [Note by AKTA:] When the ordinate $q' = 0$ the letters from left to right along the abscissa $s'$ should read as follows: $H^\circ$, $A$, $K$, $F'$ and $F$. Due to a misprint the first point $H^\circ$ was printed as $H$. When the ordinate $q'$ is equal to $-1$, the letters along the abscissa from left to right are $G$ and $G'$. Close to $q' = -1.5$ and $s' = -3$ we have $H$, while close to $q' = -8.5$ and $s' = -2.8$ we have $H'$.}

\begin {figure} [htb]
\centerline {\includegraphics[width=.9\textwidth]{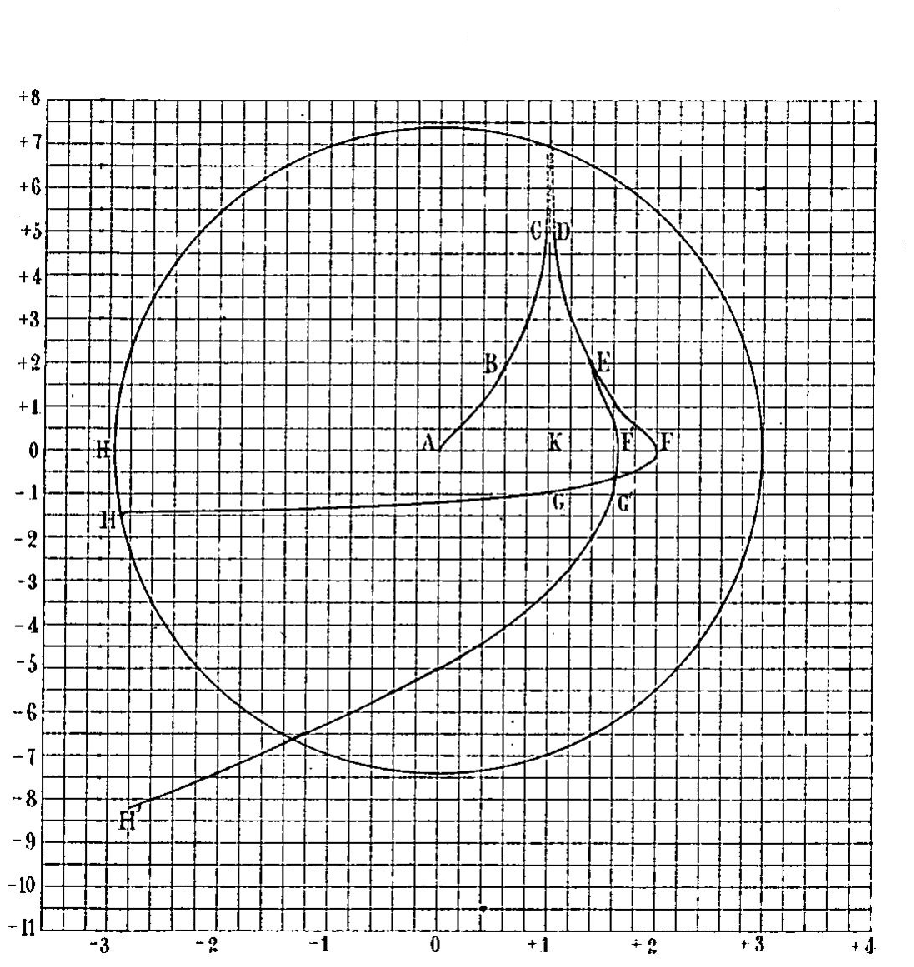}}
\caption {[Ordinate $q'$ as function of abscissa $s'$.]}
\label {fig-excerpt-EM7}
\end {figure} 

One can see from this overview that the particle $\mu$, which would have covered the distance $\frac {1}{2}g\vartheta^2$ in the time $\vartheta$ due to the acceleration $g$ coming from the {\it non-electric} force, covers twice this path under the joint action of the {\it electric} force; moreover, while it had reached the velocity $g\vartheta$ without the electric force, it now reaches an {\it infinitely large velocity} with [the joint action of] the electric force.

However, with this attained {\it infinitely large velocity}, the particle $\mu$ does not cover the {\it smallest finite path element}, due to the fact that at the same moment the {\it acceleration} $dq/dt$, which became equally infinitely large, suddenly jumps from $+\infty$ to $-\infty$, that is, changes to an infinitely large {\it deceleration}, causing the {\it velocities} to become equal long {\it before} and {\it after} this moment. For instance, the velocity $q$ at time $t = + \vartheta$ (that is, after the time interval $2\vartheta$ calculated from the beginning of the motion) is equal to the velocity in the beginning, at time $t = -\vartheta$, namely $q = 0$, where the path $s$, when the spherical shell is large enough for $s$ to still have room inside it, would have grown again by $g\vartheta^2$, so that $s$ would become $= 2g\vartheta^2$. The charge $\varepsilon'$ would thereby have grown up to $2\eta$. From now on, however, with time and charge [of the spherical shell] continuing to increase, the displacement of the particle $\mu$ from the center of the shell would decrease quickly up to $s = 0$, and then become negative up to $s = -R$, where the particle $\mu$ would hit the spherical shell at time $t$, which can be determined through the equation

\[
-R = g\vartheta^2\left[1 + \frac {t}{\vartheta}\left(1 - \frac {1}{2}\log\frac {t^2}{\vartheta^2}\right)\right]\ ,
\]
and with the velocity $q$ which, after $t$ has been determined, is found from the equation $q = [g\vartheta/2]\log[t^2/\vartheta^2]$.

It has been assumed up to now, that the radius $R$ of the sphere is larger than the largest value which $s$ has reached at time $t = +\vartheta$, namely, $2g\vartheta^2$. If $R$ were smaller, then it is evident that the particle $\mu$ would have collided earlier against the spherical shell, namely, at the moment in which $s$ would become $= R$, which can be determined from the equation

\[
R = g\vartheta^2\left[1 + \frac {t}{\vartheta}\left(1 + \frac {1}{2}\log\frac {t^2}{\vartheta^2}\right)\right]\ .
\]

Now, finally, when there is no continuous increase in the electric charge $\varepsilon'$, as previously assumed, but instead of this the charge $\varepsilon'$ remains {\it constant} after it reaches the value $\eta$ and surpasses it by any assumed arbitrarily small value, then let us designate this constant charge as $\eta(1 + e^{-n})$, and consequently the time at which this occurred as $t = +e^{-n}\vartheta$, the velocity of the particle $\mu$ at this moment as $q = ng\vartheta$, and the distance of the particle from the center of the sphere as $s = (1 + (1 + n)e^{-n})g\vartheta^2$. This results in the differential equation:

\[
dq = -\frac {ae^n}{\mu}\cdot dt\ ,
\]
and from it through integration:

\[
q = -\frac {ae^n}{\mu}t + C\ .
\]
Now if the time is calculated from the moment in which the charge [on the spherical shell] has become constant, where the velocity $q = ng\vartheta$, thus yielding $C = ng\vartheta$, therefore, as [the constant force] $a$ has been set $=g\mu$, [we obtain]:

\[
q = \frac {ds}{dt} = -ge^n\cdot t + ng\vartheta\ .
\]
From this one obtains through a second integration:

\[
s = ng\vartheta t - \frac {1}{2}ge^n\cdot t^2 + C'\ ,
\]
and, as has already been mentioned, for $t = 0$ we have the value from $s = (1 + (1 + n)e^{-n})g\vartheta^2$, yielding consequently:

\[
C' = \left(1 + (1 + n)e^{-n}\right)g\vartheta^2\ ,
\]
therefore:

\[
s = ng\vartheta\cdot t - \frac {1}{2}ge^n\cdot t^2 + \left(1 + (1 + n)e^{-n}\right)g\vartheta^2\ .
\]
This formula for the displacement $s$ and the obtained formula for the velocity, namely:

\[
q = -ge^n\cdot t + ng\vartheta
\]
are now used, {\it for a constant remaining charge} $\varepsilon'$, to determine all motions of the particle $\mu$. They can be represented in a tabular overview, for instance in the following Table for the case in which $n = 2$, when $s/(g\vartheta^2) = s'$ and $q/(g\vartheta) = q'$ are set as above:

\begin{center}
\begin{tabular}{|c|c|c|c|}
\hline
 & & & \\
$\frac{t}{\vartheta}$ & $s'$ & $q'$ & $\frac{\varepsilon'}{\eta}$\\
 & & & \\
\hline
 & & & \\
0 & $1 + \frac {6}{2e^2}$ & 2 & $1 + \frac {1}{e^2}$\\
 & & & \\
$\frac {1}{e^2}$ & $1 + \frac {9}{2e^2}$ & 1 & ---\\
 & & & \\
$\frac {2}{e^2}$ & $1 + \frac {10}{2e^2}$ & 0 & ---\\ 
& & & \\
$\frac {3}{e^2}$ & $1 + \frac {9}{2e^2}$ & $-1$ & ---\\
 & & & \\
$\frac {4}{e^2}$ & $1 + \frac {6}{2e^2}$ & $-2$ & ---\\
 & & & \\
$\frac {5}{e^2}$ & $1 + \frac {1}{2e^2}$ & $-3$ & ---\\
 & & & \\
$\frac {6}{e^2}$ & $1 - \frac {6}{2e^2}$ & $-4$ & ---\\
 & & & \\
\hline
\end {tabular}
\end {center}

This Table can easily be continued; but one can see already from it that, after the charge [on the spherical shell] has become constant, from the time $t = 2\vartheta/e^2$ onwards, the displacement of the particle $\mu$ from the center of the shell decreases and very soon becomes negative, until finally the particle $\mu$, when $s$ becomes $= - R$, collides against the spherical shell, at time $t$ and with the velocity $q$, which can be determined from the two equations:

\[
- R = \left(1 + \frac {3}{e^2}\right)g\vartheta^2 + 2g\vartheta\cdot t - \frac {e^2}{2}g\cdot t^2\ ,
\]

\[
q = 2g\vartheta - e^2g\cdot t\ .
\]

One can see from this presentation of the whole process in its {\it context}, that none of the ``inconsistent or absurd'' consequences, by which Helmholtz wanted to refute the established fundamental law, actually occur.

The curve $ABCDE$ on page \pageref {fig-excerpt-EM7} represents the dependence of the velocity $q$ as a function of the displacement $s$ of the particle $\mu$ from the center of the sphere, {\it with a uniformly increasing charge} $\varepsilon'$, up to the moment when this charge becomes greater than $\eta$, namely, $= \eta(1 + [1/e^2])$. This curve can now be continued in two ways, {\it either} for a {\it charge} [on the spherical shell] {\it continuing to grow} uniformly as before, which is represented by the curve $EFGH$ and which has already been considered, or for a charge $\varepsilon' = \eta(1 + [1/e^2])$ which remains {\it constant} from now on, which is related to the determinations in the Table mentioned above, after which the curve $EF'G'H'$ forms the continuation of curve $ABCDE$.

In both cases the particle $\mu$ moves in a continuous path, namely, in the first case along a straight line from $A$ up to $F$ and from there back to $A$ and further to $H^\circ$,\footnote {[Note by AKTA:] See Footnote \ref {foot-Ho} on page \pageref {foot-Ho}.} where the particle hits the spherical shell; in the second case along a straight line from $A$ up to $F'$ and from there back to $A$ and $H^\circ$.

Also the velocity of the particle along its path changes always continuously, except at {\it one} point $K$, in the middle of the path $AF$, where the velocity of the particle becomes infinitely large, and at the same time with it the work performed from the beginning of the motion onwards. But if we represent this performed work as {\it positive}, this is immediately followed by a {\it negative} case which is also infinitely large.

Each of these two performed works can be divided into two parts, namely, the {\it first} or {\it positive} case of the work performed along the path from $A$ to a point at a distance $= [(n + 1)/e^n]\cdot g\vartheta^2$ before $K$, and in the work performed along this last distance {\it before} $K = [(n + 1)/e^n]g\vartheta^2$; the {\it latter} or {\it negative} case of the work performed on the way through the distance {\it after} $K = [(n + 1)/e^n]g\vartheta^2$, and on the {\it rest of the way up to} $F$ or $F'$.

Of these four performed works, the two on the path $= [(n + 1)/e^n]g\vartheta^2$ {\it before} and {\it after} $K$ are {\it infinitely large}, but {\it oppositely equal}, while the other two are also oppositely equal, but have {\it finite} values. Since $n$ can now be considered so large, that the time [interval] of the first two, infinitely large performed works, namely, $2\vartheta/e^n$, can be regarded as negligible, one has two infinitely large, but oppositely equal performed works taking place in an infinitely small period of time, which, as is self-evident, have no physical effect or meaning at all.

Instead of the example above, where $n$ was $= 2$, one can choose another example, where $n$ is much larger, so that the difference of the charge $\varepsilon'$, which became constant, from $\eta$ becomes vanishingly small; no substantial change is brought about by this and one can see from the presentation of the whole process in context, that none of the ``inconsistent and absurd'' consequences, by which Helmholtz wanted to refute the established fundamental law, ever really take place.


\newpage

\begin{thebibliography}{Web78b}

\bibitem[Hel72a]{helmholtz1872b}
H.~Helmholtz.
\newblock Ueber die {T}heorie der {E}lektrodynamik.
\newblock {\em Monatsberichte der Berliner Akademie der Wissenschaften}, pages
  247--256, 1872.
\newblock Reprinted in H. Helmholtz, Wissenschaftliche Abhandlungen (Johann
  Ambrosius Barth, Leipzig, 1882), Vol. 1, Article 34, pp. 636-646.

\bibitem[Hel72b]{helmholtz1872a}
H.~von Helmholtz.
\newblock On the theory of e\-lec\-tro\-dy\-na\-mics.
\newblock {\em Philosophical Magazine}, 44:530--537, 1872.

\bibitem[Hel73]{helmholtz1873}
H.~v. Helmholtz.
\newblock Ueber die {T}heorie der {E}lektrodynamik. {Z}weite {A}bhandlung.
  {K}ritisches.
\newblock {\em Journal f\"ur die reine und angewandte Mathematik}, 75:35--66,
  1873.
\newblock Reprinted in H. Helmholtz, Wissenschaftliche Abhandlungen (Johann
  Ambrosius Barth, Leipzig, 1882), Vol. 1, Article 35, pp. 647-683; with
  additional material from 1881 on pp. 684-687.

\bibitem[Neu74]{neumann1874}
C.~Neumann.
\newblock Ueber das von {W}eber f\"ur die elektrischen {K}r\"afte aufgestellte
  {G}esetz.
\newblock {\em Abhandlungen der K\"oniglich S\"achsischen Gesellschaft der
  Wissenschaften zu Leipzig, mathematisch-physischen Classe}, 11:77--200, 1874.

\bibitem[Web46]{weber1846}
W.~Weber.
\newblock Elek\-tro\-dy\-namis\-che {M}aass\-bes\-tim\-mungen --- \"{U}\-ber
  ein all\-ge\-mei\-nes {G}rund\-ge\-setz der e\-lek\-tris\-chen {W}ir\-kung.
\newblock {\em A\-bhand\-lun\-gen bei Be\-gr\"un\-dung der K\"o\-niglich
  S\"a\-chsischen Ge\-sell\-schaft der Wis\-sen\-schaf\-ten am Ta\-ge der
  zwei\-hun\-dert\-j\"ah\-ri\-gen Ge\-burt\-stag\-fei\-er Leib\-ni\-zen's
  he\-raus\-ge\-ge\-ben von der F\"urst\-lich Ja\-blo\-nows\-kis\-chen
  Ge\-sells\-chaft (Leip\-zig)}, pages 211--378, 1846.
\newblock Reprinted in Wilhelm Weber's {\em Werke}, Vol. 3, H. Weber (ed.),
  (Springer, Berlin, 1893), pp. 25-214.

\bibitem[Web71]{weber1871}
W.~Weber.
\newblock Elek\-tro\-dy\-namis\-che {M}aass\-bes\-tim\-mun\-gen
  ins\-be\-son\-de\-re \"u\-ber das {P}rincip der {E}rhaltung der {E}nergie.
\newblock {\em Abhandlungen der K\"oniglich S\"achsischen Gesellschaft der
  Wissenschaften zu Leipzig, mathematisch-physischen Classe}, 10:1--61, 1871.
\newblock Reprinted in Wilhelm Weber's {\em Werke}, Vol. 4, H. Weber (ed.),
  (Springer, Berlin, 1894), pp. 247-299.

\bibitem[Web72]{weber1872}
W.~Weber.
\newblock Electrodynamic measurements --- {S}ixth memoir, relating specially to
  the principle of the conservation of energy.
\newblock {\em Philosophical Magazine}, 43:1--20 and 119--149, 1872.
\newblock Translated by Professor G. C. Foster, F.R.S., from the {\it
  Abhandlungen der mathem.-phys. Classe der K\"oniglich S\"achsischen
  Gesellschaft der Wissenschaften}, vol. x (January 1871).

\bibitem[Web75]{weber1875}
W.~Weber.
\newblock Ueber die {B}ewegung der {E}lektricit\"at in {K}\"orpern von
  molekularer {K}onstitution.
\newblock {\em Annalen der Physik und Chemie}, 156:1--61, 1875.
\newblock Reprinted in Wilhelm Weber's {\em Werke}, Vol. 4, H. Weber (ed.),
  (Springer, Berlin, 1894), pp. 312-357.

\bibitem[Web78a]{weber1878a}
W.~Weber.
\newblock Elek\-tro\-dy\-namis\-che {M}aass\-bes\-tim\-mun\-gen
  ins\-be\-son\-de\-re \"u\-ber die {E}nergie der {W}echselwirkung.
\newblock {\em Abhandlungen der K\"oniglich S\"achsischen Gesellschaft der
  Wissenschaften zu Leipzig, mathematisch-physischen Classe}, 11:641--696,
  1878.
\newblock Reprinted in Wilhelm Weber's {\em Werke}, Vol. 4, H. Weber (ed.),
  (Springer, Berlin, 1894), pp. 361-412.

\bibitem[Web78b]{weber1878b}
W.~Weber.
\newblock Ueber die {E}nergie der {W}echselwirkung.
\newblock {\em Annalen der Physik und Chemie}, 4:343--373, 1878.
\newblock Reprinted in Wilhelm Weber's {\em Werke}, Vol. 4, H. Weber (ed.),
  (Springer, Berlin, 1894), pp. 413-419.

\bibitem[Web87]{weber1887}
W.~Weber.
\newblock Mesures \'electrodynamiques.
\newblock In J.~Joubert, editor, {\em Collection de M\'emoires relatifs a la
  Physique, {\em Vol. III:} M\'emoires sur l'\'Electrodynamique}, pages
  289--402. Gauthier-Villars, Paris, 1887.

\bibitem[Web94]{weber1894a}
W.~Weber.
\newblock {\em Wilhelm {W}eber's Werke, {\em H. Weber, (ed.)}}, volume 4, {\em
  Galvanismus und Elektrodynamik}, second part.
\newblock Springer, Berlin, 1894.

\bibitem[Web07]{Weber2007b}
W.~Weber, 2007.
\newblock Determinations of electrodynamic measure: concerning a universal law
  of electrical action, 21st Century Science \& Technology, posted March 2007,
  translated by S. P. Johnson, edited by L. Hecht and A. K. T. Assis. Available
  at \url {http://21sci-tech.com/translation.html} and \url
  {www.ifi.unicamp.br/~assis}.

\bibitem[Web21]{weber2020c}
W.~Weber, 2021.
\newblock Electrodynamic measurements, especially on the energy of interaction.
  Fourth version posted in February 2021 at \url {www.ifi.unicamp.br/~assis}.
  Translated by Joa Weber. Edited by A. K. T. Assis.

\end{thebibliography}
\end {document}